\documentclass[aps,prl,twocolumn,groupedaddress,amsmath,superscriptaddress]{revtex4-1}
\usepackage{amsfonts}
\usepackage{amssymb}
\usepackage{amsmath}
\usepackage[dvips]{graphicx}
\usepackage{color}

\definecolor{nblue}{rgb}{0.3,0.3,1.0}
\definecolor{ngreen}{rgb}{0.2,0.7,0.2}
\definecolor{nred}{rgb}{0.9,0.1,0}
\definecolor{nblack}{rgb}{0,0,0}

\usepackage{amsmath}
\usepackage{graphicx}
\usepackage{amsfonts}
\usepackage{amssymb}
\usepackage{hyperref}
\usepackage{color}
\usepackage{mathrsfs}
\usepackage{isomath}
\usepackage{amsmath}
\usepackage{amsthm}
\usepackage{epstopdf}
\usepackage{txfonts}
\newcommand{\an}[1]{\left\langle{#1}\right\rangle}

\newcommand{\ket}[1]{\left|#1 \right\rangle}

\newcommand{\Exp}[1]{\langle #1\rangle}

\definecolor{maroon}{rgb}{0.7,0,0}

\begin{document}
	
	\title{Experimentally ruling out joint reality based on operational completeness}
	
	\author{Qiuxin Zhang$^{\ddagger}$}
	\address{Department of Physics, Renmin University of China, Beijing 100872, China}
	\author{Yu~Xiang$^{\ddagger}$}
	\address{State Key Laboratory for Mesoscopic Physics, School of Physics, Frontiers Science Center for Nano-optoelectronics, $\&$ Collaborative Innovation Center of Quantum Matter, Peking University, Beijing 100871, China}
	\author{Xiaoting~Gao}
	\address{State Key Laboratory for Mesoscopic Physics, School of Physics, Frontiers Science Center for Nano-optoelectronics, $\&$ Collaborative Innovation Center of Quantum Matter, Peking University, Beijing 100871, China}
	\author{Chenhao Zhu}
	\address{Department of Physics, Renmin University of China, Beijing 100872, China}
	\author{Yuxin Wang}
	\address{Department of Physics, Renmin University of China, Beijing 100872, China}
	\author{Liangyu Ding}
	\address{Department of Physics, Renmin University of China, Beijing 100872, China}
	\author{Xiang Zhang}
	\address{Department of Physics, Renmin University of China, Beijing 100872, China}
	\address{Beijing Academy of Quantum Information Sciences, Beijing 100193, China}
	\author{Shuaining Zhang}
	\email{zhangshuaining@ruc.edu.cn}
	\affiliation{Department of Physics, Renmin University of China, Beijing 100872, China}
	\affiliation{Beijing Academy of Quantum Information Sciences, Beijing 100193, China}
	\author{Shuming Cheng}
	\address{The Department of Control Science and Engineering $\&$ Shanghai Institute of Intelligent Science and Technology, Tongji University, Shanghai 201804, China}
	\address{Institute for Advanced Study, Tongji University, Shanghai, 200092, China}
	\author{Michael J. W. Hall}
	\address{Department of Theoretical Physics, Research School of Physics, Australian National University, Canberra, Australian Capital Territory 0200, Australia}
	\author{Qiongyi~He}
	\address{State Key Laboratory for Mesoscopic Physics, School of Physics, Frontiers Science Center for Nano-optoelectronics, $\&$ Collaborative Innovation Center of Quantum Matter, Peking University, Beijing 100871, China}
	\address{Collaborative Innovation Center of Extreme Optics, Shanxi University, Taiyuan, Shanxi 030006, China}
	\author{Wei Zhang}
	\email{wzhangl@ruc.edu.cn}
	\address{Department of Physics, Renmin University of China, Beijing 100872, China}
	\address{Beijing Academy of Quantum Information Sciences, Beijing 100193, China}
	
	\begin{abstract}		
	Whether the observables of a physical system admit real values is of fundamental importance to a deep understanding of nature. In this work, we report a device-independent experiment to confirm that the joint reality of two observables on a single two-level system is incompatible with the assumption of operational completeness, which is strictly weaker than that of preparation noncontextuality. We implement two observables on a trapped $^{171}{\rm Yb}^{+}$ ion to test this incompatibility via violation of certain inequalities derived from both linear and nonlinear criteria. Moreover, by introducing a highly controllable dephasing channel, we show that the nonlinear criterion is more robust against noise. Our results push the fundamental limit to delineate the quantum-classical boundary and pave the way for exploring relevant problems in other scenarios.
	\end{abstract}
	
	\maketitle
	
%%%%%%%%%%%
	\textit{Introduction.--}  The physical reality of an observable can be identified with the existence of a predetermined value that is revealed if the corresponding measurement procedure is implemented on a physical system. The question of whether quantum observables  have physical reality was raised by Einstein, Podolsky, and Rosen in 1935~\cite{EPR35}, and has attracted great interest from theoretical and experimental sides. While joint reality is interpretationally consistent for commuting observables~\cite{modal}, a variety of {\it no-go theorems} have been formulated that exclude joint reality of incompatible observables under various assumptions. Distinctive examples include Bell inequalities~\cite{Bell65,BrunnerRMP} and Kochen-Specker (KS) theorems~\cite{ks,contextreview}, derived under respective assumptions of locality and measurement noncontextuality, and various no-go theorems have been verified in many physical experiments~\cite{bellexp1,bellexp2,bellexp3,bellexp4,contextnature,context1,context2,Mikenc,Mikeprxq,Pryde09,Howardcontext,Wang22}. These results enrich our understanding of nature at the conceptual level, and also find wide practical applications, including cryptographic tasks~\cite{Ekert91} and randomness generation~\cite{Pironio10}.

	 It is particularly desirable to rule out the joint reality of any {\it two} incompatible quantum observables, under the weakest possible assumptions. Yet most no-go theorems only rule out the simultaneous reality of three or more observables (and typically only for quantum systems with three or more dimensions). This includes Bell inequalities~\cite{Bell65,BrunnerRMP,bellexp1,bellexp2,bellexp3,bellexp4} and Kochen-Specker theorems~\cite{ks,contextreview,contextnature,context1,kcbs,Cabellokcbs,Yu12,context2,Wang22}, as well as no-go results based on steering~\cite{Jevtic15} and generalized measurement noncontextuality, where measurements are operationally equivalent to have the same probability distribution of outcomes in every quantum/ontic state~\cite{Spekkens05, Mikenc,Howardcontext,Howardcontext2,Cabelloprl}. 	
	 Moreover,  existing results applicable to just two observables~\cite{Pryde09,PuseyPRA,Mikeprxq} require an assumption of preparation noncontextuality~\cite{Spekkens05}, i.e., two state preparations that are operationally equivalent correspond to the same mixture over ontic states in the ontological model, and have only been tested under additional assumptions of parity obliviousness~\cite{Pryde09} and/or measurement noncontextuality~\cite{Mikeprxq}. Thus, a fundamental question naturally arises: Can the joint reality of two observables be tested under less stringent assumptions?

	Here, we address this question by experimentally demonstrating that an assumption of {\it operational completeness}  is sufficient to rule out joint reality of two incompatible observables,  even for  a single qubit. Operation completeness is a strictly weaker assumption than preparation noncontextuality, stating that if two ensembles are operationally similar, i.e., the statistics of all measurable observables are approximately identical up to errors from finite statistics, the joint relative frequencies of any two observables with predetermined real values must approximately the same~\cite{MichaelPRA}. Using a single trapped $^{171}{\rm Yb}^{+}$ ion, we perform two device-independent (DI) no-go tests in the form of linear and nonlinear inequalities~\cite{MichaelPRA}. With the techniques of microwave operation and highly distinguishable state-dependent fluorescence detection, we achieve quantum gates with very high fidelity ($99.97\%$) and low measurement error ($2.08\%$). It is observed that both inequalities are  significantly violated, thus conclusively ruling out the possibility that two incompatible observables can have real preexisting values under the assumption of operational completeness. Further, we test the noise robustness of the two tests by introducing a controllable dephasing channel (of some interest in its own right), and conclude that the nonlinear criterion is more robust against noise. The controllability of dephasing in a quantitative level is first realized in trapped ion systems, and can extend the scope of its application in the study of quantum systems with decay.

%%%%%%%%%%%%%%%%%

\begin{figure}[tbp]
		\centering
		\includegraphics[width=\linewidth]{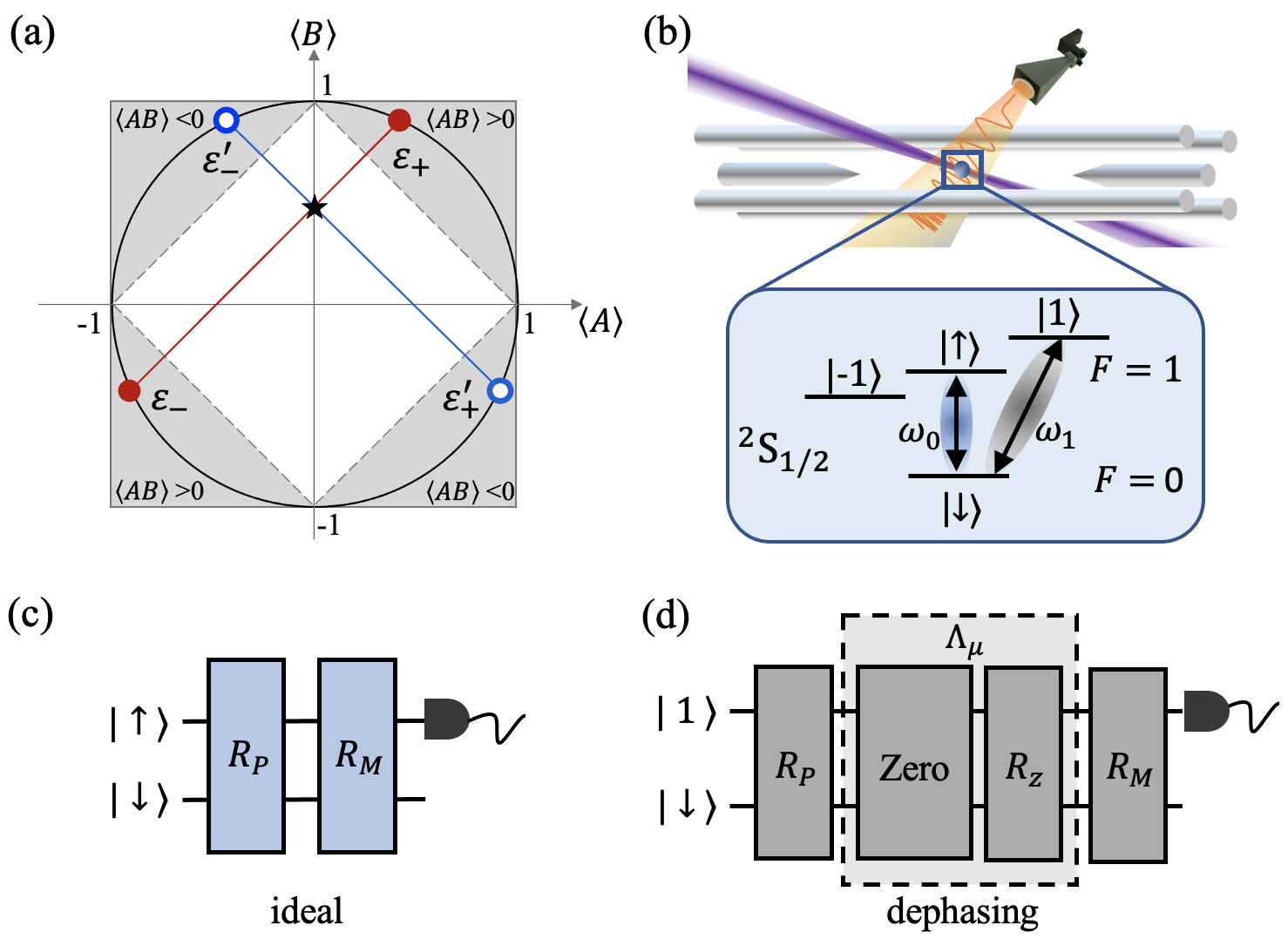}
		\caption{
			(a) Geometry underlying no-go result~(\ref{the3}) for $c=0$. The red and blue dots depict four ensembles that violate the inequality, corresponding to the qubit ensembles in Eq.~(\ref{StateSets}). 
			(b) The four-rod Paul trap used in experiment and the level diagram of $^{171}$Yb$^+$. The purple beam is the 369.5nm laser for Doppler cooling, optical pumping and state detection. Microwaves are generated from a microwave horn antenna.
			(c) The experimental sequence for preparation and measurement of the ideal (no dephasing) states. Preparation (P) and measurement (M) operations are both implemented by single qubit gates $R_{P (M)}(\Theta,\phi)$, where $\Theta$ and $\phi$ are controlled by driven signals. 
			(d) The experimental sequence for the dephasing case. The dephasing process $\Lambda_\mu$ comprises a decoherence period labeled by ``Zero" and a $z$-axis rotation $R_z(\Theta)=R_x^{-1}(\pi/2)R_y(\Theta)R_x(\pi/2)$. }
		\label{fig1}
	\end{figure}

\textit{No-go inequalities.--} Consider two observables $A$ and $B$ of a physical system each having two measurement outcomes, labelled as $\pm 1$. They are said to have joint reality if the corresponding measurement procedures reveal definite predetermined outcomes $\alpha, \beta=\pm 1$~\cite{endnote}. For an ensemble of $N$ systems, we define  the average expectation $\an{AB}=\sum_{\alpha,\beta}\alpha\beta N(\alpha,\beta)/N$ where $N(\alpha,\beta)$ denotes the number of systems having real values $A=\alpha$ and $B=\beta$. Note that the range of $\an{AB}$ is constrained by the observed averages $\an{A}=\sum_{\alpha,\beta}\alpha N(\alpha,\beta)/N$ and $\an{B}=\sum_{\alpha,\beta}\beta N(\alpha,\beta)/N$. In particular, substituting $\alpha, \beta=\pm 1$ into the relation $N(\alpha,\beta)/N=[1+\alpha\Exp{A}+\beta\Exp{B}+\alpha\beta\Exp{AB}]/4\geq 0$ yields~\cite{MichaelPRA}
%	\begin{equation}
%		\begin{aligned}
%			&\Exp{AB}>c, \quad \text{for}~\Exp{A}+\Exp{B}>1+c~\text{or}~\Exp{A}+\Exp{B}<-1-c,\\
%			&\Exp{AB}<c, \quad \text{for}~\Exp{A}-\Exp{B}>1-c~\text{or}~\Exp{A}-\Exp{B}<-1+c,
%		\end{aligned}\label{jr}
%	\end{equation}
\begin{equation}
\begin{aligned}
&\Exp{AB}>c \quad \text{for}\quad |\Exp{A}+\Exp{B}|>1+c,\\
&\Exp{AB}<c \quad \text{for}\quad |\Exp{A}-\Exp{B}|>1-c,
\end{aligned}\label{jr}
\end{equation}
for any $c\in (-1, 1)$. These constraints on  $\an{AB}$  are illustrated as shaded regions in Fig.~\ref{fig1}(a), for the choice $c=0$ (which is optimal for orthogonal qubit observables~\cite{MichaelPRA}).

	A geometric no-go argument follows by combining the joint reality constraints~(\ref{jr}) with operational completeness. First, as depicted in Fig.~\ref{fig1}(a) for $c=0$, if two ensembles $\varepsilon_{+}$ and $\varepsilon_{-}$ of $N$ systems satisfy $\an{AB}_{\varepsilon_{\pm}}>c$ (solid red dots), then $\an{AB}_{\varepsilon}>c$ for any mixture $\varepsilon$ of $\varepsilon_{+}$ and $\varepsilon_{-}$ (lying on the red solid line). Similarly, any mixture $\varepsilon'$ of two ensembles $\varepsilon'_{+}$ and $\varepsilon'_{-}$ that satisfy $\an{AB}_{\varepsilon'_{\pm}}<c$ (hollow blue dots) must have $\an{AB}_{\varepsilon'}<c$. Hence, at the intersection point (black star) of the two lines we have the strict inequality $\an{AB}_{\varepsilon'}<c<\an{AB}_{\varepsilon}$.

 	Further, if the mixtures $\varepsilon$ and $\varepsilon'$ at the intersection point are operationally equivalent (e.g., via direct test, or because the ensembles $\varepsilon_{\pm}, \varepsilon'_{\pm}$ lie in an operational plane of $A$ and $B$ so that any two convex combinations with the same values of $\langle A\rangle$ and $\langle B\rangle$ are operationally equivalent~\cite{MichaelPRA}), 
 	%Further, if the ensembles $\varepsilon_{\pm}, \varepsilon'_{\pm}$ lie in an operational plane of $A$ and $B$, i.e., if any two convex combinations having the same values of $\langle A\rangle$ and $\langle B\rangle$ are operationally equivalent~\cite{MichaelPRA}, then the mixtures $\varepsilon$ and $\varepsilon'$ at the intersection point must be operationally equivalent. Operational 
 	operational completeness then implies the predetermined real values of $A$ and $B$ have equal joint relative frequencies up to statistical errors, i.e.,  $N(\alpha,\beta|\varepsilon)/N\approx N(\alpha,\beta|\varepsilon')/N$. Thus, $\an{AB}_{\varepsilon}\approx \an{AB}_{\varepsilon'}$, which contradicts the strict inequality above. Hence, the existence of such ensembles is incompatible with joint reality under the assumption of operational completeness. It is remarked that this assumption is strictly weaker than preparation noncontextuality used in~\cite{Spekkens05, Pryde09, PuseyPRA, Mikenc,Mikeprxq}, and that the argument is device-independent in the sense that it requires no description of the preparation and measurement devices.

	Finally, it follows via the relations in Eq.~(\ref{jr}) that one can obtain a simple sufficient criterion for joint reality~\cite{MichaelPRA}, in the form of the linear inequality
\begin{equation}
		\begin{aligned}
			\ell_c=&\min\bigg\{\Exp{A}_{\varepsilon_{+}}+\Exp{B}_{\varepsilon_{+}}-1-c, -\Exp{A}_{\varepsilon_{-}}-\Exp{B}_{\varepsilon_{-}}-1-c,\\
			&\Exp{A}_{\varepsilon'_{+}}-\Exp{B}_{\varepsilon'_{+}}-1+c, -\Exp{A}_{\varepsilon'_{-}}+\Exp{B}_{\varepsilon'_{-}}-1+c\bigg\}\leq 0
		\end{aligned}\label{the3}
	\end{equation}
	for all $c \in (-1,1)$. For $c=0$ this corresponds to at least one of the four ensembles $\{ \varepsilon_{\pm}, \varepsilon'_{\pm} \}$ lying in the white region of Fig.~\ref{fig1}(a), rather than in  a shaded region. A nonlinear necessary and sufficient criterion may also be obtained~\cite{MichaelPRA}:
	\begin{equation}
		\tau=\left|
		\begin{array}{cccc}
			\Exp{A}_{\varepsilon_{+}}&\Exp{B}_{\varepsilon_{+}}&p\Exp{A}_{\varepsilon_{+}}+q\Exp{B}_{\varepsilon_{+}}-1&1\\ \Exp{A}_{\varepsilon'_{+}}&\Exp{B}_{\varepsilon'_{+}}&r\Exp{A}_{\varepsilon'_{+}}+s\Exp{B}_{\varepsilon'_{+}}+1&1\\
			\Exp{A}_{\varepsilon'_{-}}&\Exp{B}_{\varepsilon'_{-}}&-r\Exp{A}_{\varepsilon'_{-}}-s\Exp{B}_{\varepsilon'_{-}}+1&1\\ \Exp{A}_{\varepsilon_{-}}&\Exp{B}_{\varepsilon_{-}}&-p\Exp{A}_{\varepsilon_{-}}-q\Exp{B}_{\varepsilon_{-}}-1&1
		\end{array}
		\right|\leq 0\label{the4}
	\end{equation}
	for all $p,q,r,s=\pm1$ with $pqrs=-1$ (generalizing the preparation noncontextuality result in~\cite{PuseyPRA}). Quantum realizations can violate the above two inequalities, up to $\ell_{\text{max}}=\sqrt{2}-1\approx 0.414$ and $\tau_{\text{max}}=8(\sqrt{2}-1)\approx 3.314$ respectively, by measuring  observables $A=\sigma_x$ and $B=\sigma_z$ on the  qubit  ensembles described by Bloch vectors  $\textbf{\textit{n}}_{\varepsilon_{\pm}}=\pm(1,0,1)/\sqrt{2}$ %$\textbf{\textit{n}}_{\varepsilon_{-}}=(-1,0,-1)/\sqrt{2}$,
	and $\textbf{\textit{n}}_{\varepsilon'_{\pm}}=\pm(1,0,-1)/\sqrt{2}$.  
	%, and $\textbf{\textit{n}}_{\varepsilon'_{-}}=(-1,0,1)/\sqrt{2}$. 

	%
	
%%%%%%%%%%%%%%%	
	\textit{Robustness against noise.--} The maximal violations $\ell_{\rm max}$ and $\tau_{\rm max}$ are achieved for a specific choice of observables and a perfect preparation of ensembles. It is then of great importance to study to what extent the quantum violation of these reality tests still pertains under experimental imperfections, such as state transmission loss and detection error. For this purpose, we consider a practical scenario where the ensembles before measurement suffer dephasing noise, which can be modeled as the quantum channel
	\begin{equation}
		\Lambda_{\mu} (\rho)=\left(1-\frac{\mu}{2}\right) \rho+\frac{\mu}{2} \sigma_z \,\rho \,\sigma_z.
		\label{dephaseCh}
	\end{equation}
Here, $\rho$ describes the density operator and the parameter $\mu \in [0, 1] $ is a dephasing factor which shrinks state ensembles in the $x$-$y$ plane. For example, if $\mu=1$ all qubit states in the $x$-$y$ plane undergo complete dephasing and reduce to a classical bit $\{0,1\}$; and if $\mu=0$ the dephasing is absent and the state ensembles remain unchanged.
	
	We fix the observables as $A =\sigma_x$ and $B =\sigma_z$ and consider a suitable one-parameter set of state ensembles
	\begin{equation} 
		\begin{aligned}
			\textbf{\textit{n}}_{\varepsilon_{+}}&=(\sqrt{1-m^2_+},0,m_+),~\textbf{\textit{n}}_{\varepsilon_{-}}=(-\sqrt{1-m^2_-},0,m_-), \\
			\textbf{\textit{n}}_{\varepsilon'_{+}}&=(\sqrt{1-m^2_-},0,m_-),~\textbf{\textit{n}}_{\varepsilon'_{-}}=(-\sqrt{1-m^2_+},0,m_+), \label{StateSets}
		\end{aligned}
	\end{equation}
	where $m_{\pm}=(\cos\theta\pm \delta_\theta)/2$ with $\delta_\theta=\sqrt{2- \cos^2\theta}$ are dependent on a parametrized angle $\theta \in [0, \pi/2]$. The maximal violation for the two inequalities (\ref{the3}) and (\ref{the4}) under dephasing $\mu$ is achieved for $c=0$, yielding
	\begin{equation} 
		\begin{aligned}
			\ell_{\text{max}}^\theta(\mu)&=  -\mu m_++\delta_\theta-1, \\ 
			\tau_{\text{max}}^\theta (\mu)&= 4\delta_\theta(1-\mu)\left[ -\mu\sin^2\theta+\delta_\theta(\delta_\theta-1)\right].
		\end{aligned}
	\end{equation}
 In our experiment, we generate these ensembles by setting $\theta=\pi/3$, as depicted by the red and blue dots in Fig.~\ref{fig1}(a). Under the ideal condition with no dephasing, this set of ensembles yield the violations $\ell^\theta_{\text{max}}=(\sqrt{7}-2)/2 \approx 0.323 $ and $\tau^\theta_{\text{max}}=7(\sqrt{7}-2)/2 \approx 2.260 $. With increasing dephasing factor $\mu$, the nonlinear inequality Eq.~(\ref{the4}) is more robust in the sense that $\tau_{\text{max}}^\theta (\mu)$ can still be positive to witness the incompatibility while the corresponding $\ell_{\text{max}}^\theta(\mu)$ becoming negative, which confirms the conjecture made in~\cite{MichaelPRA}. (See Supplemental Materials~\cite{supp} for more details.)
	
%%%%%%%%%%%%%%%	
	\textit{Experimental setup.--}  The setup is composed of a single $^{171}{\rm Yb}^{+}$ ion trapped in a four-rod Paul trap, as shown in Fig.~\ref{fig1}(b). The hyperfine levels in the $^{2}{\rm S}_{1/2}$ ground-state manifold, $\ket{\downarrow}\equiv\ket{F=0,m_F=0}$ and $\ket{\uparrow}\equiv\ket{F=1,m_F=0}$, are encoded as the computation basis of a single ion-qubit. The transition frequency between these two clock states is $\omega_{0}= 2\pi \times 12.64281$ GHz. This ion-qubit is initialized to the state $\ket{\downarrow}$ via $500 \mu s$ Doppler cooling and $2.7 \mu s$ optical pumping, and measured by shining a $370$nm laser to excite the state $\ket{\uparrow}$ to the $^2$P$_{1/2}$ manifold and then collecting fluorescent photons with a photo-multiplier tube~\cite{Olmschenk}. The $\pi$ time of the Rabi oscillation is $5.89\mu$s, leading to the Rabi frequency $\Omega_{\ket{\downarrow}, \ket{\uparrow}}=2\pi \times84.9$ kHz. The error of detection a $\ket{\uparrow}$ state population for $\ket{\downarrow}$ is $1.71\%$, while it is $2.08\%$ to detect $\ket{\downarrow}$ for the state $\ket{\uparrow}$. These detection errors are further calibrated via the method proposed in~\cite{Detection}. 
	
	To prepare the four state ensembles on the operational plane corresponding to the $x$-$z$ plane, as per Eq.~(\ref{StateSets}), we apply microwaves with the corresponding resonant frequency on the ion-qubit. By tuning the duration and phase of the microwaves we can realize a general class of single-qubit rotations
	\begin{equation}
		R(\Theta,\phi)=
		\left(
		\begin{array}{ccc}
			\cos(\Theta/2)& -i e^{-i\phi}\sin(\Theta/2)\\
			-i e^{i\phi}\sin(\Theta/2)&\cos(\Theta/2)
		\end{array}
		\right) ,
		\label{rotation}
	\end{equation}   
	where $\Theta$ and $\phi$ are the rotation angle and phase, respectively. The experimental gate sequence is shown in Fig.~\ref{fig1}(c), and the parameters to prepare the four state ensembles $\{ \varepsilon_+, \varepsilon_-, \varepsilon'_+,  \varepsilon'_- \}$
%$\{\mathcal{P}_{\varepsilon_{+}},~\mathcal{P}_{\varepsilon_{-}},~\mathcal{P}_{\varepsilon'_{+}},~\mathcal{P}_{\varepsilon'_{-}}\}$  [MH: Why not $\varepsilon_\pm, \varepsilon'_\pm$?]  
are respectively 
		$$\begin{matrix} 
			R_{P}(0.134967\pi,\pi/2), & R_{P}(0.634967\pi,3\pi/2),\\ R_{P}(0.634967\pi,\pi/2), &  R_{P}(0.134967\pi,3\pi/2).
     	\end{matrix}$$
	
To realize the noise process described by Eq.~(\ref{dephaseCh}), we introduce the Zeeman state $\ket{F=1,m_F=1}$ as state $\ket{1}$,  which is more sensitive to the decoherence process, and couple it to the ground state $\ket{\downarrow}$. It can be regarded as a noisy qubit with transition frequency $\omega_{1}=\omega_{0}+2\pi \times9.6$ MHz, as shown in Fig.~\ref{fig1}(b). The Rabi frequency is adjusted to $\Omega_{\ket{\downarrow},\ket{1}} = 2\pi \times 181.2$ kHz. {Interestingly, by analyzing the dynamics of decoherence induced by magnetic field fluctuation, we find that the dephasing process is consistent with the decoherence process with an additional $z$-axis rotation $R_z(\Theta)$. The $z$-axis rotation $R_z$ is included to ensure the prepared dephasing states to lie in the $x$-$z$ plane. The decoherence process is labeled by ``Zero" since the amplitude of microwave is set to zero during that period. For a specific dephasing factor $\mu$, the duration of ``Zero" is determined by $\mu  T $, where $T$ is the coherence time of our system. This provides a feasible scheme to precisely control the dephasing channel, where the dephasing factor $\mu$ in Eq.~(\ref{dephaseCh}) can be tuned by controlling the evolution time of the noisy qubit interacting with the environment~\cite{supp}.
In the experiment, the dephasing factor is tuned and estimated at four typical values of $\mu = 0.014, 0.220, 0.512$, and $0.653$ via operation sequences depicted in Figs.~\ref{fig1}(c) and \ref{fig1}(d).
	
	\begin{figure}[tbp]
		\centering
		\includegraphics[width=\linewidth]{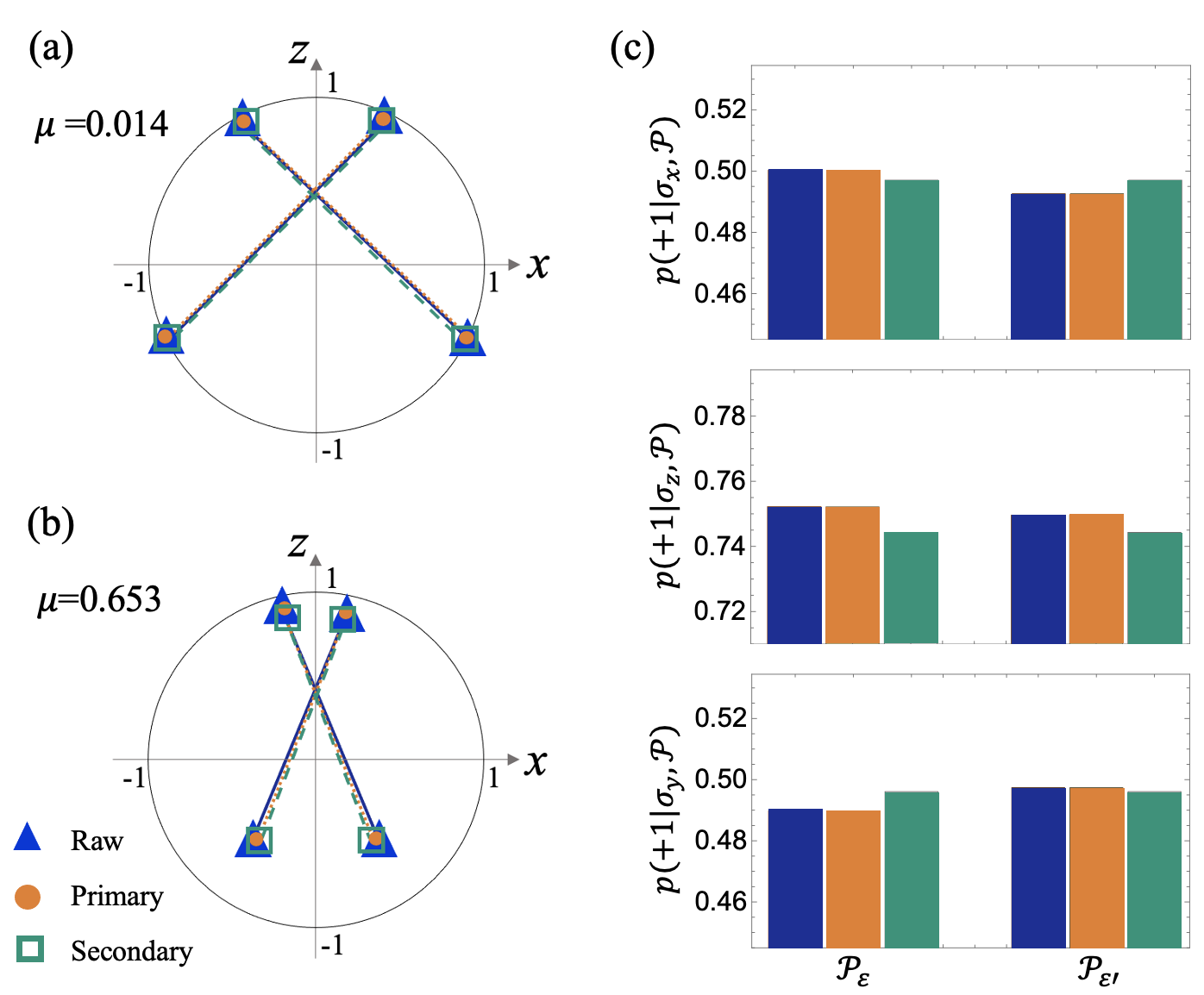}
		\caption{The results of data analysis for operational equivalence. The states prepared in experiments, $|\psi\rangle_{\varepsilon_{+}}$,~ $|\psi\rangle_{\varepsilon_-}$,~$|\psi\rangle_{\varepsilon^{\prime}_{+}}$ and $|\psi\rangle_{\varepsilon^{\prime}_-}$, are plotted on the $x$-$z$ plane of the Bloch sphere with dephasing factors (a)~$\mu = 0.014$ and (b)~$\mu=0.653$. Raw preparations (blue triangles) are given from the experimental data. Primary preparations (orange dots) are chosen from the region encircled by raw data to fit GPT theory. Secondary preparations (green rectangle) are chosen to ensure operational equivalence. (c) Probabilities of getting $+1$ outcome from the primary measurements of $\sigma_x$ (top), $\sigma_z$ (middle), and $\sigma_y$ (bottom) on raw preparations (left blue bars), primary preparations (middle orange) and secondary preparations (right green), with $\mu$=0.014. }
		\label{fig2}
	\end{figure}
	
	It is remarked that, although the dephasing channel in the form of Eq.~(\ref{dephaseCh}) does not distort the $x$-$z$ plane of the Bloch sphere where the four states~(\ref{StateSets}) reside, in realistic experimental conditions the prepared states may not be perfectly constrained in the $x$-$z$ plane and thus would be non-homogeneously affected under dephasing. Hence, we need to add the two eigenstates of $\sigma_y$ as an extra dimension to correct this preparation inaccuracy. This calibration procedure is introduced in the subsequent discussion and details are given in the Supplemental Materials~\cite{supp}. To ensure the measurements for the set of preparations~(\ref{StateSets}) are tomographically complete, we also implement the observable $\sigma_y$, as have been done to calibrate the choice of preparation and measurement observables and suggested in Refs.~\cite{PuseyPRA,Mikenc,MichaelPRA,Mikeprxq}. 
	
	The observable $\sigma_z$ can be directly measured by state-dependent fluorescence detection in the $\sigma_z$ basis, while the measurements of $\sigma_x$ and $\sigma_y$ are realized by first applying a $\pi/2$ rotation along the $y$- and $x$-axis denoted as $R_{M}(\pi/2,\pi/2)$ and $R_{M}(\pi/2,0)$, respectively. We collect the outcome statistics  for many trials and record the relative frequencies as desired. Here, we stress that although measurements are taken for all three spin axes in experiments, the raw data are rearranged into a form as measured from two binary-outcome observables based on two calibrated spin axes when testing the no-go inequalities~\cite{supp}.
	
%%%%%%%%%%%%%%%
	\textit{Data analysis and results.--} 
	A great advantage of the no-go inequalities Eqs.~(\ref{the3}) and (\ref{the4}) is that only local relative frequencies $p(k|j,\mathcal{P})$ are required, which is defined as the probability that an outcome $k\in\{-1, +1\}$ is obtained by measuring an observable $j\in\{\sigma_x,~\sigma_z\}$ on the state ensemble obtained by state preparation $\mathcal{P}$. In our experiment, the state preparations satisfy $\mathcal{P}_{\varepsilon}=t\mathcal{P}_{\varepsilon_+}+(1-t)\mathcal{P}_{\varepsilon_-}$ and $\mathcal{P}_{\varepsilon'}=(1-t)\mathcal{P}_{\varepsilon'_+}+t\mathcal{P}_{\varepsilon'_-}$ with $t=(\sqrt{7}+1)/2\sqrt{7}\approx 0.689$. We run the state preparation and measurement sequence 6890 times for $\textbf{\textit{n}}_{\varepsilon_{+}}$ ($\textbf{\textit{n}}_{\varepsilon'_{-}}$) and 3110 times for $\textbf{\textit{n}}_{\varepsilon_{-}}$ ($\textbf{\textit{n}}_{\varepsilon'_{+}}$)  to realize the ensembles mixture. The fidelity of state preparation can be obtained from standard quantum state tomography. For the nearly ideal case with dephasing factor $\mu = 0.014$, the average fidelity of the four state ensembles is $99.94\%\pm 0.05\%$, while for the cases with $\mu = 0.220, 0.512,$ and $0.653$, the average fidelities read $99.74\%\pm 0.24\%$, $99.89\%\pm 0.17\%$, and $99.89\%\pm 0.08\%$, respectively.

	We then check whether the two state preparations $\mathcal{P}_{\varepsilon}$ and $\mathcal{P}_{\varepsilon'}$ are operationally equivalent, i.e., $p(k|j,\mathcal{P}_{\varepsilon})\approx p(k|j,\mathcal{P}_{\varepsilon'})$ for all $j, k$. To establish a device-independent test, we also have to avoid presupposing that our experiment data is well fit by a quantum description. Instead, the physical state preparations and measurements must be fit into the more abstract framework of generalized probabilistic theories (GPT)~\cite{Mikenc,Mikeprxq}. In our experiment, we first fit the raw data of actual state preparations $\mathcal{P}^\text{r}$ (blue triangles in Figs.~\ref{fig2}(a) and \ref{fig2}(b) for different dephasing cases) to a set of generalized states where three binary-outcome measurements are tomographically complete~\cite{supp}, leading to the estimations of the primary state preparations $\mathcal{P}^{\text{p}}$ (orange dots) as shown in Figs.~\ref{fig2}(a) and \ref{fig2}(b). 
	
However, due to the error of quantum operations and measurements, the outcome probabilities  $p(k|j,\mathcal{P}^\text{p})$ under primary procedures $\mathcal{P}^{\text{p}}$ in general cannot meet the requirement of operational equivalence, i.e., $p(k|j,\mathcal{P}_{\varepsilon}^{\text{p})}\neq p(k|j,\mathcal{P}_{\varepsilon'}^{\text{p})}$ for all $j$ and $k$. Therefore, we further infer a $4\times4$ secondary probability matrix $\mathcal{S}$ whose elements $\mathcal{S}_{ln}=p(+1|j_n,\mathcal{P}_l^\text{s})$ with $\mathcal{P}^\text{s}=(\mathcal{P}_{\varepsilon_{+}}^\text{s},~\mathcal{P}_{\varepsilon_{-}}^\text{s},~\mathcal{P}_{\varepsilon'_{+}}^\text{s},~\mathcal{P}_{\varepsilon'_{-}}^\text{s})^{\top}$. Each secondary preparation $\mathcal{P}_l^\text{s}$ can be defined by a probabilistic mixture of the six primaries (two additional primaries along the $y$-axis for calibration). By minimizing their displacement from the primaries while ensuring the secondary preparations meet the operational equivalence assumption constraint, the secondary probabilities can be obtained (see details in Supplemental Materials~\cite{supp}). Notice that the correction induced by the secondary procedure would change the prepared states and consequently the dephasing factor. Specifically, if the primary state preparations (orange dots in Fig.~\ref{fig2}(a)) are ideal without dephasing, the secondary ones (green rectangles) would in general acquire a small but nevertheless finite dephasing factor, such that the system is only {\it nearly} ideal. Results with other values of the dephasing factor are shown in Supplemental Materials~\cite{supp}. Using the techniques described above, we report the probability of obtaining outcome ``+1" for each measurement-preparation pair and verify the requisite operational equivalence, as shown in Fig.~\ref{fig2}(c).
	
\begin{figure}[tbp]
	\centering
	\includegraphics[width=0.9\linewidth]{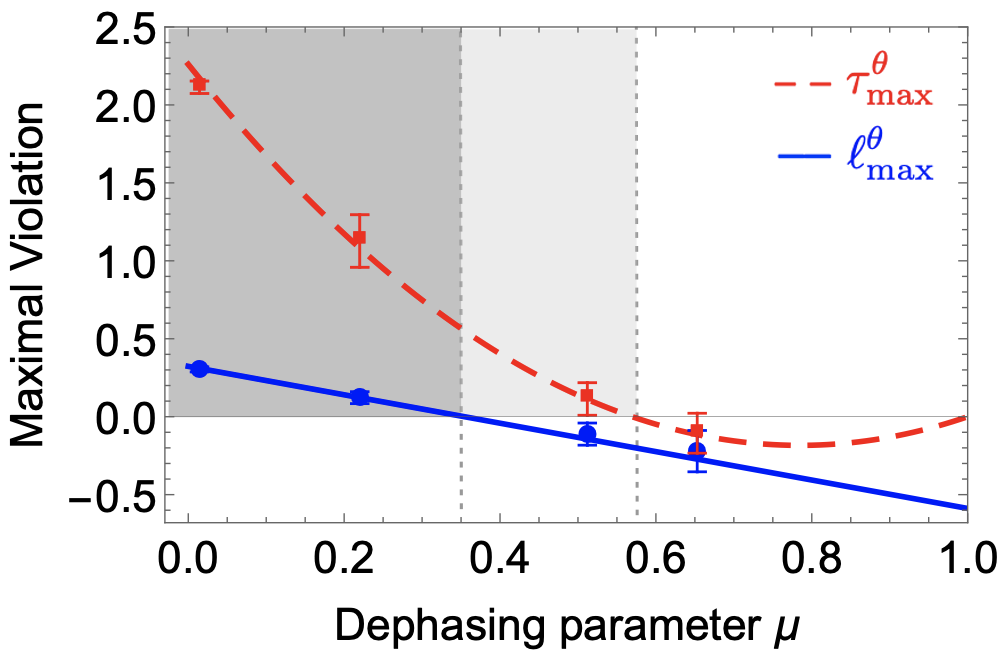}
	\caption{The maximum violation of no-go inequalities (\ref{the3}) (lower blue) and (\ref{the4}) (upper red) for varying values of dephasing factor $\mu$. The ensembles are prepared in forms of Eq.~(\ref{StateSets}) with $\theta = \pi/3$. The experimental data (solid dots) are in good agreement with theoretical predictions (solid and dashed curves). The error bars represent one standard deviation and are obtained based on the statistics of finite sampling in experiment. The theoretical thresholds for the two inequalities are $\mu=3-\sqrt{7} \approx 0.354$ and $\mu=(7-2\sqrt{7})/3 \approx 0.569$, respectively, which are depicted by vertical dotted lines.} % [MH: Put $\tau^\theta_{\max}$ first in legend?]  
	\label{fig3}
\end{figure}

	The central results of the no-go theorem tests are shown in Fig.~\ref{fig3}, where the maximal violations of inequalities~(\ref{the3}) and (\ref{the4}) are plotted as functions of the dephasing factor $\mu$. One can clearly observe that Eq.~(\ref{the3}) is weaker than Eq.~(\ref{the4}), and is more fragile to state dephasing. Specifically, linear inequality~(\ref{the3}) is violated only in the relatively narrow range $\mu<3-\sqrt{7} \approx 0.354$, while nonlinear inequality~(\ref{the4}) remains valid up to $\mu < (7-2\sqrt{7})/3 \approx 0.569$. This difference can be understood by noting that the geometric argument for the linear inequality requires all four state ensembles to lie in respective regions constrained by Eq.~(\ref{jr}), e.g., the four shaded regions in Fig.~\ref{fig1}(a). This is fulfilled only for $\mu<3-\sqrt{7}$. In contrast, the derivation of the nonlinear inequality relaxes this requirement, by also taking into account the weights of the four ensembles required for the mixtures $\varepsilon$ and $\varepsilon'$ to be operationally equivalent, corresponding to the intersection point in Fig.~\ref{fig1}(a)~\cite{MichaelPRA}. Thus, in the parameter regime $\mu \in (3-\sqrt{7}, (7-2\sqrt{7})/3)$, despite two of the four state ensembles crossing from the shaded regions into the white region of Fig.~\ref{fig1}(a) [see Fig.~\ref{fig2}(b)], the nonlinear inequality is still able to rule out the joint reality of the two observables.

	The main error sources in our experiment are the gate infidelity ($0.03\%$) caused by the fluctuation of microwave parameters, and the dephasing channel error (0.12\%) resulting from the time-dependent decoherence process and the uncertainty of estimated dephasing factor $\mu$. The former is characterized by  the  randomized benchmarking method~\cite{RB} and the latter is estimated by the infidelity of  the  experiment-prepared dephasing state~\cite{supp}. For the nearly ideal case with $\mu = 0.014$, we find the similarity between secondary preparations and their corresponding primaries can reach $99.06\%$ averaging over all runs. Here, the value of similarity is defined by the weights in the mixture of four secondaries~\cite{supp}. For the other three cases with larger dephasing factor, high similarities ($ \ge 91.48\%$) are also achieved. 

	\textit{Conclusion.}-- We report two device-independent experimental no-go tests of joint reality of two observables with a single trapped ion. Our results show that in the simple but nontrivial single qubit system, joint reality is inconsistent with an assumption of operational completeness, which is strictly weaker than those found in literature. With high fidelities of state preparation, gate operation, and measurement, we observe significant violations for both testing inequalities, and find perfect agreement with theoretical predictions. Moreover, we investigate the noise effect on the two inequalities by introducing a controllable dephasing channel, and conclude that the inequality with non-linear form is more robust against dephasing than the linear one. Our work not only deepens the understanding of quantum reality,  but also provides a scheme to explore noise effects in quantum applications or even to take advantage of noise in certain circumstances, such as error-mitigating schemes~\cite{zhang-EM} in noisy intermediate scale quantum information processing.
	
\begin{acknowledgments}
{\it Acknowledgements.--} This work is supported by the Beijing Natural Science Foundation (Grant No. Z180013, Z190005), the National Natural Science Foundation of China (Grant No. 91836106, 11975026, 12074427, 12074428, 12004011, 12125402), the Shanghai Municipal Science and Technology Fundamental Project (Grant No. 21JC1405400), the National Key R\&D Program of China (Grant No. 2018YFA0306501), and the China Postdoctoral Science Foundation (Grant No. BX20200379, 2021M693478).
\newline
$^{\ddagger}$ Q. Zhang and Y. Xiang contributed equally to this work.
	
\end{acknowledgments}
	
%%%%%%%%%%%%%%%%%%


\begin{thebibliography}{99}
\bibitem{endnote} Note that while joint reality implies the existence of joint statistics for observables $A$ and $B$, there is no assumption made of measurement compatibility~\cite{comp1,comp2,comp3,comp4}, i.e, of the existence of a joint measurement of $A$ and $B$ with these statistics.
\end{thebibliography}
\end{document}